\RequirePackage{fix-cm}
\documentclass[twocolumn]{svjour3}          
\smartqed  
\pdfoutput=1
\usepackage{graphicx}
\usepackage{mathptmx}      
\usepackage{subfigure}
\usepackage{amsmath}
\usepackage{tikz}
\usepackage{pgfplots}
\usepackage{natbib}

\journalname{Microfluidics and Nanofluidics}
\begin{document}

\title{Three dimensional simulations of viscous folding in diverging microchannels}


\author{Bingrui XU         \and
        Jalel Chergui      \and
        Seungwon Shin      \and
        Damir Juric
}


\institute{Bingrui XU \at
              Laboratoire FAST, Univ. Paris-Sud, CNRS, Universit\'e Paris-Saclay, F-91405 Orsay, France \\
              LIMSI, CNRS, Universit\'e Paris-Saclay, F-91405 Orsay, France \\
              \email{xu@fast.u-psud.fr}
           \and
           Damir Juric \and Jalel Chergui\at
              LIMSI, CNRS, Universit\'e Paris-Saclay, F-91405 Orsay, France \\
              Damir Juric \\
              \email{damir.juric@limsi.fr}\\
              Jalel Chergui \\
              \email{jalel.chergui@limsi.fr}
           \and
           Seungwon Shin \at
              Department of Mechanical and System Design Engineering, Hongik University, Seoul 121-791, Republic of Korea\\
           }

\date{Received: date / Accepted: date}

\maketitle

\begin{abstract}
Three dimensional simulations on the viscous folding in diverging microchannels reported by \citet{Cubaud2006afolding} are performed using the parallel code BLUE for multi-phase flows \citep{shin2014solver}. The more viscous liquid $L_1$ is injected into the channel from the center inlet, and the less viscous liquid $L_2$ from two side inlets. Liquid $L_1$ takes the form of a thin filament due to hydrodynamic focusing in the long channel that leads to the diverging region. The thread then becomes unstable to a folding instability, due to the longitudinal compressive stress  applied to it by the diverging flow of liquid $L_2$. Given the long computation time, we were limited to a parameter study comprising five simulations in which the flow rate ratio, the viscosity ratio, the Reynolds number, and the shape of the channel were varied relative to a reference model. In our simulations, the cross section of the thread produced by focusing is elliptical rather than circular. The initial folding axis can be either parallel or perpendicular to the narrow dimension of the chamber. In the former case, the folding slowly transforms via twisting to perpendicular folding , or it may remain parallel. The direction of folding onset is determined by the velocity profile and the elliptical shape of the thread cross section in the channel that feeds the diverging part of the cell. Due to the high viscosity contrast and very low Reynolds numbers, direct numerical simulations of this two-phase flow are very challenging and to our knowledge these are the first three-dimensional direct parallel numerical simulations of viscous threads in microchannels.  Our simulations provide good qualitative comparison of the early time onset of the folding instability however since the computational time for these simulations is quite long, especially for such viscous threads, long-time comparisons with experiments for quantities such as folding amplitude and frequency are limited.
\keywords{Three dimensional \and viscous folding \and diverging microchannel}
\end{abstract}

\section{Introduction}
\label{intro}
The folding of viscous threads in diverging microchannels has recently attracted much attention due to the need to mix two fluids with very different viscosities. The dynamics of viscous multiphase flows at small scales is important in industrial technology (oil recovery, biodiesel production, etc.). Microfluidic devices are well suited for studying precisely controlled flow geometries and finely manipulating the fluid, and can be used to produce individual bubbles, droplets and complex soft materials \citep{utada2005monodisperse,cubaud2005bubble,meleson2004formation}. The effective mixing is of great importance in these various microfluidic applications. But microfluidic flows are usually laminar, so liquid streams are parallel and different fluids can only mix by diffusion. The time scale associated with diffusion, $t_{d}=h^2/D$, where $h$ is the characteristic length scale and $D$ is the diffusion coefficient between the liquids, is typically much larger than the time scale associated with convection, $t_c=h/U$, where $U$ is the characteristic flow velocity. Therefore, diffusion alone is an extremely inefficient mixing method.

There are different innovative strategies to enhance mixing in microfluidics, which can be classified as either active or passive methods. In active methods an external forcing is imposed by e.g. rotary pumps \citep{chou2001microfabricated}, forced oscillatory transverse flows \citep{bottausci2004mixing} or electric or magnetic fields \citep{paik2003rapid,paik2003electrowetting,pollack2002electrowetting,kang2007achaotic,kang2007bchaotic,rida2004manipulation}. Passive methods rely on a particular design of the microchannel, including patterned surface relief \citep{chen2009performance,bringer2004microfluidic,stroock2002chaotic,stroock2002patterning}. However, industrial and biological fluids usually exhibit widely different viscosities and  the relative motions between the fluids are complex. In this article we study one promising method, wherein periodic folding of viscous threads injected into microchannels enhances mixing by greatly increasing the specific surface area of the fluid/fluid interface. 

The buckling (folding or coiling) of slender viscous threads is familiar to anyone who has ever poured honey or molten chocolate onto toast. \citet{taylor1969instability} investigated the viscous buckling problem and suggested that the instability requires an axial compressive stress, like the more familiar `Euler' buckling of a compressed elastic rod. Since then, viscous buckling has been
studied by numerous authors using experimental, theoretical, and numerical approaches \citep{cruickshank1982viscous,cruickshank1982energy,cruickshank1983theoretical,cruickshank1988low,griffiths1988folding,tchavdarov1993buckling,mahadevan1998fluid,skorobogatiy2000folding,tome1999numerical,ribe2004coiling,ribe2006stability,maleki2004liquid,habibi2014liquid}. The primary  result of this work is that buckling can occur in four distinct modes (viscous, gravitational, inertio-gravitational, and inertial) depending on the force that balances the viscous resistance to bending as a function of fall height.  

With the exception of \citet{griffiths1988folding}, all the studies cited above consider `non-immersed' folding/coiling that occurs when the influence of the external fluid (typically air in experiments) is negligible. Recently, \citet{Cubaud2006afolding} have studied the immersed buckling
that occurs when two fluids with different viscosities are injected into a diverging microchannel. The thread is produced by hydrodynamic focusing of a viscous fluid flow by a less viscous fluid injected
from the sides. Silicone oils with different viscosities were used to obtain different viscosity ratios. On the basis of their experimental results, \citet{Cubaud2006afolding} proposed that 
$f \sim \dot{\gamma}$, where $f$ is the folding frequency and $\dot{\gamma}=U_1/(h/2)$ is the characteristic shear rate.  The thread of radius $R_1$ can be assumed to flow at nearly constant velocity, $U_1=Q_1/(\pi R_1^2)$ , like a solid plug, inside a sheath of the less viscous liquid, similar to the flow in a
circular channel. In this case, $U_1$ represents the maximum velocity of the surrounding liquid. Downstream, the thread and surrounding liquid enter the diverging channel creating a decelerating extensional flow. Extensional viscous stresses cause the thread to bend and fold, rather than dilate, in order to minimize dissipation and conserve mass. As the thread folds, it
reduces its velocity and mixes with the outer liquid. In addition to folding, many other potentially useful flow phenomena are obtained, including oscillatory folding, folding modified by strong diffusion, heterogeneous folding, and subfolding \citep{cubaud2006bfolding}.

\citet{chung2010numerical} performed numerical and experimental studies on viscous folding in diverging microchannels similar to those of \citet{Cubaud2006afolding}. However, it is important to note that the numerical simulations of \citet{chung2010numerical} are two-dimensional, unlike their or Cubaud's experiments which are fully three-dimensional. \citet{chung2010numerical} obtained a regime diagram for the flow pattern observed (stable, folding, or chaotic) as a function of the flow rate ratio, the viscosity ratio, and the channel shape. In addition to  the divergence angles $\alpha=\pi/2$ and $\alpha=\pi$, \citet{chung2010numerical} also performed simulations for a channel with walls of hyperbolic shape, to obtain a more uniform compressive stress along the channel's centerline. The hyperbolic channel generated folding flows with smaller frequency and amplitude, as well as a delay of onset of the folding. There are two main differences between Chung's simulations and Cubaud's experiments. First, \citet{chung2010numerical} found the existence of an upper bound of viscosity ratio for folding instability. Secondly, \citet{chung2010numerical} obtained a power-law relation $f \sim \dot{\gamma}^{1.68}$, which is quite different from the \citet{Cubaud2006afolding} law $f \sim \dot{\gamma}$.

For fluids with a large viscosity ratio, the thread generated by hydrodynamic focusing requires a long focusing channel to become thin. The existence of an upper bound of viscosity ratio for the folding instability could be owed to the thick thread, because the focusing channel is short in the study of \citet{chung2010numerical}.
Three-dimensional simulations promise to be helpful in understanding such details of the viscous folding phenomenon. To this end, we use the parallel code BLUE for multiphase flow based on the front tracking method (developed by \citet*{shin2014solver}) to simulate three-dimensional viscous folding in diverging microchannels. The computational domain includes a focusing channel sufficiently long to allow the full formation of threads as in the experiments of \citet{Cubaud2006afolding}.

\section{Methods}
\label{sec:methods}

\subsection{Mathematical formulation}
\label{sec:math-form}

Here we describe the basic solution procedure for the Navier-Stokes
equations with a brief explanation of the interface method. The governing
equations for an incompressible two-phase flow can be expressed by a single field formulation as follows: 
\begin{align}
  \label{eq:nav-con} &\nabla \cdot \boldsymbol{u}=0,\\
  \label{eq:nav-mom} &\rho \Big( \frac{\partial \boldsymbol{u}}{\partial
    t}+\boldsymbol{u} \cdot \nabla\boldsymbol{u}\Big)=-\nabla P+\rho
  \boldsymbol{g}+\nabla \cdot \mu(\nabla\boldsymbol{u}+\nabla
  \boldsymbol{u}^T)+\boldsymbol{F}.
\end{align}
where $\boldsymbol{u}$ is the velocity, $P$ is the pressure, $\boldsymbol{g}$
is the gravitational acceleration, and $\boldsymbol{F}$ is the local surface
tension force at the interface. 

The fluid properties such density or viscosity are defined in the entire computational domain:
\begin{align}
  \label{eq:den-do} \rho(\boldsymbol{x},t)&=\rho_1+(\rho_2-\rho_1)I(\boldsymbol{x},t),\\
  \label{eq:vis-do} \mu(\boldsymbol{x},t)&=\mu_1+(\mu_2-\mu_1)I(\boldsymbol{x},t).  
\end{align}
Where the subscripts 1 and 2 stand for the respective phases. The indicator function $I(\boldsymbol{x},t)$,  a numerical Heaviside function, is zero in one phase and unity in the other phase. Numerically, $I$ is resolved with a sharp but smooth transition across 3 to 4 grid cells and is generated using a vector distance function computed directly from the tracked interface \citep{shin2009hybrid, shin2009simulation}.

The surface tension $\boldsymbol{F}$ can be described by the hybrid formulation
\begin{equation}
  \label{eq:surface-tension}
  \boldsymbol{F}=\sigma \kappa_H \nabla I,
\end{equation}
where $\sigma$ is the surface tension coefficient, $\kappa_H$ is twice the mean interface curvature field calculated on the Eulerian grid using
\begin{align}
  \label{eq:inter-curvature}
  &\kappa_H=\frac{\boldsymbol{F}_L \cdot
    \boldsymbol{G}}{\sigma\boldsymbol{G} \cdot \boldsymbol{G}},\\
  &\boldsymbol{F}_L=\int_{\Gamma(t)} \sigma \kappa_f\boldsymbol{n}_f
  \delta_f(\boldsymbol{x}-\boldsymbol{x}_f) \mathrm{d}s,\\
  &\boldsymbol{G}=\int_{\Gamma(t)} \boldsymbol{n}_f
  \delta_f(\boldsymbol{x}-\boldsymbol{x}_f) \mathrm{d}s.
\end{align}
Here $\boldsymbol{x}_f$ is a parameterization of the interface $\Gamma(t)$,
and $\delta(t)$ is a Dirac distribution that is non-zero only when
$\boldsymbol{x}=\boldsymbol{x}_f$. $\boldsymbol{n}_f$ is the unit normal
vector to the interface and $\mathrm{d}s$ is the length of the interface
element. $\kappa_f$ is again twice the mean interface curvature, but obtained
from the Lagrangian interface structure. The geometric information, unit
normal $\boldsymbol{n}_f$ and length of the interface element
$\mathrm{d}s$ in $\boldsymbol{G}$, $\boldsymbol{F}$ are computed directly from the Lagrangian interface and then distributed onto an Eulerian grid using the discrete delta function. The details following Peskin's \citep{peskin1977numerical} well known immersed boundary approach and a description of our procedure for calculating the force $\boldsymbol{F}$ and constructing the function field $\boldsymbol{G}$ can be found in \citet{shin2007high}.

The Lagrangian elements of the interface are advected by integrating
\begin{equation}
  \label{eq:Lagran-coor}
  \frac{\mathrm{d}\boldsymbol{x}_f}{\mathrm{d}t}=\boldsymbol{V}
\end{equation}
with a second order Runge-Kutta method where the interface velocity
$\boldsymbol{V} $ is interpolated from the Eulerian velocity.

\subsection{Numerical method}
\label{sec:numerical-method}

The treatment of the free surface uses the Level Contour Reconstruction Method (LCRM), a hybrid Front Tracking/Level Set technique. The LCRM retains the usual features of classic Front Tracking: to represent the interface with a triangular surface element mesh, to calculate the surface tension and advect it. A major advantage of the LCRM, compared with standard Front Tracking, is that all the
interfacial elements are implicitly instead of logically connected. The LCRM
periodically reconstructs the interface elements using a distance
function field, such as the one in the Level Set method, thus
allowing an automatic treatment of interface element restructuring and
topology changes without the need for logical connectivity between interface
elements.

The Navier-Stokes solver computes the primary variables of velocity $\boldsymbol{u}$ and pressure $P$ on a fixed and uniform Eulerian mesh by means of Chorin’s projection method
\citep{chorin1968numerical} with implicit solution of velocity.  For the spatial discretization we use the well-known staggered mesh, MAC method \citep{harlow1965numerical}. The pressure and the distance function are located at cell centers while the components of velocity are located at cell
faces. All spatial derivatives are approximated by standard second-order
centered differences.

The code structure consists essentially of two main modules: (1) a module for solution of the incompressible Navier-Stokes equations and (2)
a module for the interface solution including tracking the phase front,
initialization and reconstruction of the interface when necessary. The parallelization of the code is based on algebraic domain decomposition, where the
velocity field is solved by a parallel generalized minimum residual (GMRES) method for the implicit viscous terms and the pressure by a parallel multigrid method motivated by the algorithm of \citet{kwak2003multigrid}. Communication across process threads is handled by message passing interface (MPI) procedures.

Further detailed informations can be found in \citet{shin2014solver}.

\subsection{Problem definition}
\label{sec:problem-definition}

\begin{figure}[!t]
  \centering
  \includegraphics[width=0.48\textwidth]{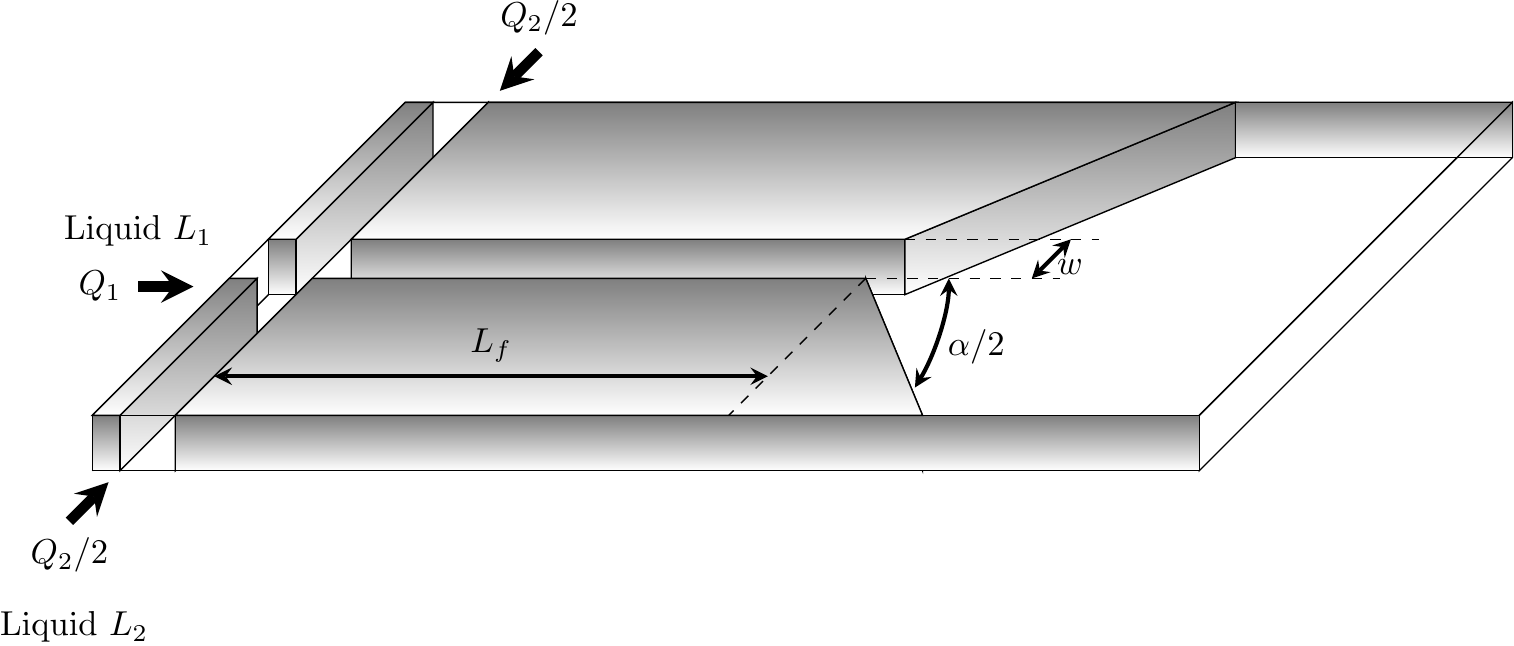}
\caption{The calculation domain of the microchannel. The width of
the inlets and the microchannel is $w$ and the length of the
focusing microchannel is $L_f$.}
\label{fig:1}       
\end{figure}
Fig. \ref{fig:1} shows the computational domain, which is similar to that
used by Cubaud and Mason \citep{Cubaud2006afolding}. The geometry is composed of two subdomains: the flow-focusing part and the flow-diverging part. The more viscous liquid $L_1$ with viscosity $\eta_1$ is injected into
the channel from the center inlet at a volumetric rate $Q_1$, and the less viscous liquid $L_2$ with
viscosity $\eta_2$ from two side inlets at a total volumetric rate $Q_2$. The interfacial tension between two liquids is $\gamma$.The dimensions of the simulation domain are $2$ mm $\times 0.25$ mm $\times 5$mm. The width of the inlets and the microchannel is $w=0.25$ mm and the length of the long focusing channel is $L_f=10w=2.5$ mm. We use a Neumann boundary condition on the outlet, where the velocity derivatives are set to $0$. 

Some important dimensionless numbers are defined as follows:
\begin{align}
  \label{eq:5-1} &\chi=\frac{\eta_1}{\eta_2},\\
  \label{eq:5-2} &\phi=\frac{Q_1}{Q_2},\\
  \label{eq:5-3} &\mathrm{Re}_1=\frac{\rho_1LV_1}{\eta_1},\\
  \label{eq:5-4} &\mathrm{Re}_2=\frac{\rho_2LV_2}{\eta_2},\\
  \label{eq:5-5} &\mathrm{Ca}_1=\frac{\eta_1 \bar{V}}{\gamma},\\
  \label{eq:5-6} &\mathrm{Ca}_2=\frac{\eta_2 \bar{V}}{\gamma}.
\end{align}
The characteristic length scale $L=0.5w$, and the characteristic velocities in Reynolds numbers $V_1$ and $V_2$ are the average velocities and can be calculated from the volume flow flux and geometry parameters $V_1=Q_1/w^2$ and $V_2=0.5Q_2/w^2$. The capillary numbers are calculated in the long focusing channel, the characteristic velocity is $\bar{V}=(Q_1+Q_2)/w^2$. Furthermore, we designed different channel geometries with two different diverging angles $\alpha=\pi$ and $\alpha=\pi/2$ for the main chamber. 

A reference simulation (case 1) is chosen and its detailed parameters and dimensionless numbers are shown in Table \ref{tab:1}.
\begin{table}[h!]
  \centering
  \caption{Dimensional and nondimensional parameters for the simulation case 1 with $\chi=2174$, $\phi=1/12$ and $\alpha=\pi/2$.}
  \label{tab:1}
  \begin{tabular}{l l l}
    \hline\noalign{\smallskip}
    Variables&Units&Values\\ 
    \noalign{\smallskip}\hline\noalign{\smallskip}
    $\rho$&kg/mm$^3$ & $0.8 \times 10^{-6}$ \\
    $\eta_1$ &kg/mm/s & $4864.28\times 10^{-6}$\\
    $\eta_2$&kg/mm/s & $2.24\times 10^{-6}$\\
    $Q_1$&mm$^3$/s &0.83333\\
    $Q_2$&mm$^3$/s &10\\
    $\gamma$&kg/s$^2$&$2.55\times 10^{-3}$\\
    $\mathrm{Re}_1$& & $2.74\times 10^{-4}$\\
    $\mathrm{Re}_2$& & 3.57\\
    $\mathrm{Ca}_1$& & 330.64\\
    $\mathrm{Ca}_2$& & 0.15\\
    $\phi$& & 1/12\\
    $\chi$& & 2174\\
    $\alpha$& & $\pi/2$\\
    \noalign{\smallskip}\hline
  \end{tabular}
\end{table}
In our parameter study, five simulations are performed. The dimensionless quantities for these cases are given in Table \ref{tab:2}. In all 5 simulations the capillary number $\mathrm{Ca}_1$ is kept constant at $330.64$, the surface tension force is small compared to the viscous force for the liquid $L_1$. All the simulations are implemented using 64 ($4\times 2 \times 8$) computational cores (subdomains) in parallel, and for each subdomain we use a $64 \times 32 \times 64$ mesh resolution. Thus the global mesh resolution for the domain is $256 \times 64 \times 512$.
\begin{table}[h!]
  \centering
  \caption{Dimensional and  nondimensional parameters for the 5 simulations}
  \label{tab:2}
  \begin{tabular}{l l l l l}
    \hline\noalign{\smallskip}
    cases & $\mathrm{Re}_1$& $\phi$ &$\chi$&$\alpha$\\
    \noalign{\smallskip}\hline\noalign{\smallskip}
    1 (reference)& $2.74\times 10^{-4}$& 1/12& 2174 & $\pi/2$\\
    2&   $1.64\times 10^{-3}$&1/12& 2174 & $\pi/2$\\
    3& $2.74\times 10^{-4}$& 1/12& 1000 & $\pi/2$\\
    4& $2.74\times 10^{-4}$& 1/5& 2174 & $\pi/2$\\
    5& $2.74\times 10^{-4}$& 1/12& 2174 & $\pi$\\
    \noalign{\smallskip}\hline
  \end{tabular}
\end{table}

\section{Numerical results}
\label{sec:numerical-results}

\subsection{Thread formation}
\label{sec:thread-formation}

In our simulations, the threads are produced by hydrodynamic focusing. The liquid is injected from a central channel, and flows that ensheath the liquid are introduced from side channels. Downstream from the junction, the fluids flow side by side, and the width and location of the stream can be controlled through the injection flow rates. The hydrodynamic focusing technique provides an effective means of controlling the passage of chemical reagent or bio-samples through microfluidic channels and has given rise to many studies aimed at understanding its physical mechanisms. Various flow-geometry relationships have been studied to create different effects, including the influence of the channel aspect ratio \citep{lee2006hydrodynamic}, the injection geometry for detaching the central stream from the walls \citep{simonnet2005two,chang2007three}, the fluid driving mechanisms \citep{stiles2005hydrodynamic} and the effect of small and moderate viscosity contrasts between the fluids  \citep{wu2005hydrodynamic}.

\begin{figure}[!t]
  \centering
  \subfigure[$z=4$ mm]{
    \includegraphics[width=0.22\textwidth]{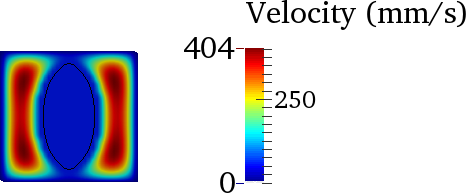}
    \label{fig:2a}
  }
  \subfigure[$z=3$ mm]{
    \includegraphics[width=0.22\textwidth]{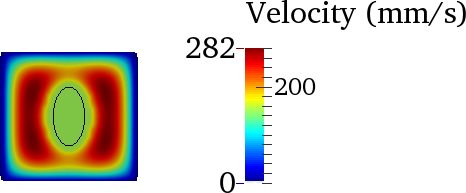}
    \label{fig:2b}
  }
  \subfigure[$z=2.5$ mm]{
    \includegraphics[width=0.22\textwidth]{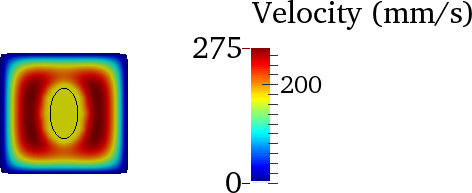}
    \label{fig:2c}
  }
  \subfigure[$z=2.2$ mm]{
    \includegraphics[width=0.22\textwidth]{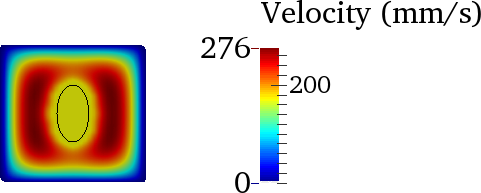}
    \label{fig:2d}
  }
\caption{The velocity contour of cross sections across the depth at different positions $z=2.2$ mm, $z=2.5$ mm, $z=3$ mm, $z=4$ mm for case 2 (Re$_1=1.64\times10^{-3}, \phi=1/12, \chi=2174, \alpha=\pi/2$), the black line is the thread interface.}
\label{fig:2}       
\end{figure}
The more viscous liquid $L_1$ passes the junction and begins to detach from the top and bottom walls. The irregular shapes on the thread near the
inlet are due to graphical artefacts. The contact line has a `V'-like shape
which is strongly stretched at the bottom. After the detachment from the walls, the liquid $L_1$ becomes thinner to form a thread. To analyse the focusing process more clearly, four cross sections across the depth at different positions $z = 2.2$ mm, $z = 2.5$ mm, $z = 3$ mm and $z = 4$ mm (case 2) are shown in Fig. \ref{fig:2}. The liquid L1 flows at an almost uniform velocity (plug flow) at the beginning of hydrodynamic focusing and is accelerated by the side flow. The thread becomes thinner (from Fig. \ref{fig:2a} to Fig. \ref{fig:2c}) and then is nearly stable (from Fig. \ref{fig:2c} to Fig. \ref{fig:2d}). Similarly, from $z = 4$ mm to $z = 2.2$ mm, the velocity contour changes dramatically at first, then slowly and at last becomes almost stable. Moreover, the flow of the liquid $L_1$ in the long microchannel is a plug flow and ensheathed by liquid $L_2$ .

\begin{figure}[!t]
  \centering
  \includegraphics[width=0.45\textwidth]{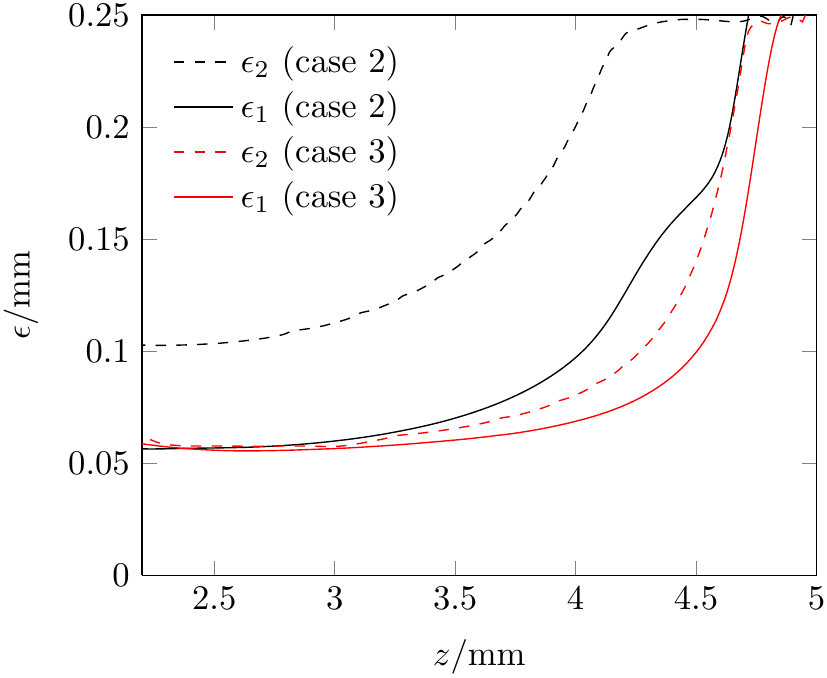}
\caption{The minor axis $\epsilon_1$ and major axis $\epsilon_2$ of the thread along the flow direction for case 2 (Re$_1=1.64\times10^{-3}, \phi=1/12, \chi=2174, \alpha=\pi/2$) and case 3 (Re$_1=2.74\times10^{-4}, \phi=1/12, \chi=1000, \alpha=\pi/2$)}
\label{fig:3}       
\end{figure}

It is noted that  the cross section of the thread is an ellipse rather than a circle. The minor axis of the thread $\epsilon_1$ and the major axis of the thread $\epsilon_2$ along the flow direction up to the diverging point are plotted in Fig. \ref{fig:3}. From Fig. \ref{fig:3}, the stable minor axis and major axis of the thread produced by focusing are $\epsilon_1 = 0.0565$ mm and $\epsilon_2 = 0.103$ mm for case 2. Besides, it suggests that liquid $L_1$ detaches completely from the walls near the position $z = 4.2$ mm. The minor axis and major axis of the thread as well as the ratios $\epsilon_1 /w$, $\epsilon_2 /w$ and $\epsilon_1 / \epsilon_2$ for all 5 cases are listed in Table \ref{tab:3}.
\begin{table}[h!]
  \caption{The minor and major axes $\epsilon_1$ and $\epsilon_2$, their ratios $\epsilon_1/\epsilon_2$, and ratios with the inlet width $\epsilon_1/w$ and $\epsilon_2/w$ of the stable thread produced by focusing for all 5 cases}
  \label{tab:3}
  \centering
  \begin{tabular}{l l l l l l l}
    \hline\noalign{\smallskip}
    cases & $\epsilon_1$& $\epsilon_2$ &$\epsilon_1/w$&$\epsilon_2/w$&$\epsilon_1/\epsilon_2$&\\
    \noalign{\smallskip}\hline\noalign{\smallskip}
    1 (base case)& 0.0573& 0.0836& 0.2292 & 0.3344&0.69\\
    2&   0.0565&0.103&0.226&0.412&0.55\\
    3& 0.0578& 0.057& 0.2312& 0.228&1.01\\
    4& 0.088& 0.111&  0.352& 0.444& 0.793\\
    5& 0.06& 0.088& 0.24 & 0.352&0.73\\
    \noalign{\smallskip}\hline
  \end{tabular}
\end{table}

\begin{figure*}[!t]
  \centering
  \subfigure[]{
    \includegraphics[width=0.45\textwidth]{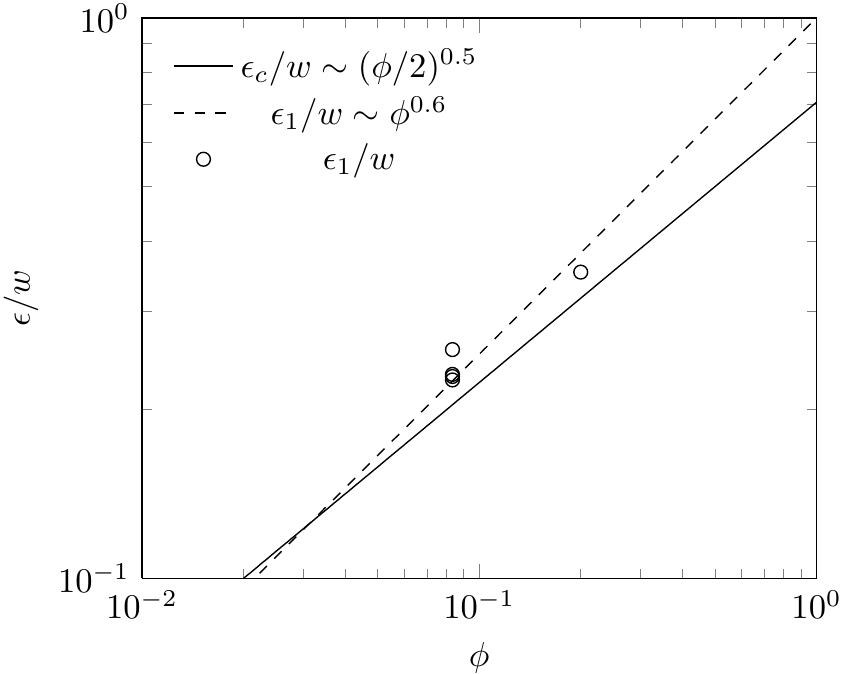}
    \label{fig:4a}
  }
  \subfigure[]{
    \includegraphics[width=0.45\textwidth]{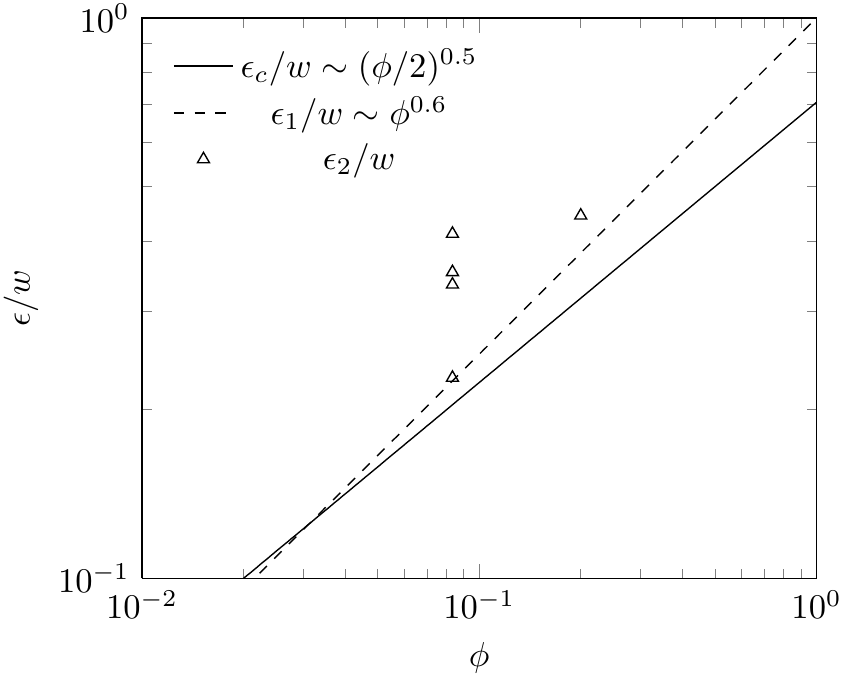}
    \label{fig:4b}
  }
\caption{The ratio $\epsilon/w$ versus flow rate
ratio $\phi$ for a thread in plug flow in a square microchannel. The solid
and dashed lines are the power-law predictions, the circle and triangle marks are our simulation results.}
\label{fig:4}       
\end{figure*}

According to the velocity profile for the annular flow in a circular tube of diameter $R$ directly calculated from the Stokes equations \citep{joseph2013fundamentals}, a simple scaling for the thread can be found with small threads $\epsilon_c/w \ll 1$ and large viscosity ratios $\chi^{-1} \ll 1$: $\epsilon_c/R \sim (\phi/2)^{0.5}$. Although this analysis is only valid for a circular tube, Cubaud \citep{cubaud2009high} suggests that the relationship between $\epsilon_c/w$ and $\phi$ is essentially the same for square tubes when $\chi^{-1} \ll 1$ and scales $\epsilon_c/w \sim (\phi/2)^{0.5}$ as for small threads. For the case of a square micro channel of width $w$, $\epsilon_c /w$ for comparing circular diameter and square cross section instead of $\epsilon_c /R$ is used. It is noted that the thread is assumed circular, so $\epsilon_c$ is used as the diameter of the thread. Cubaud’s experiments \citep{Cubaud2006afolding} suggest that the thread minor-axis (diameter) $\epsilon_1 /w$ was independent of $\chi$ and follows $\epsilon_1 /w \sim \phi^{0.6}$ . In Cubaud's experiments they took photos from
above with a high speed camera, so that only the minor-axis (diameter)
$\epsilon_1$ of the thread could be measured (Thus it is not clear whether the
thread cross section was circular or not).  In Fig. \ref{fig:4}, the estimating lines $\epsilon_c /w \sim (\phi/2)^{0.5}$, $\epsilon_1 /w \sim \phi^{0.6}$ and values $\epsilon_1 /w$, $\epsilon_2 /w$ from our simulations
are presented. When $\phi$ is small, the two power-law predictions are close.
From Fig. \ref{fig:4a}, the slope of $\epsilon_1/w$ from our simulation results agrees well with both power-law relationships $\epsilon_c /w \sim (\phi/2)^{0.5}$ and $\epsilon_1 /w \sim \phi^{0.6}$, the minor axis (the thread width from the top view) is only dependent on the flow ratio $\phi$. However for the major
axis $\epsilon_2$ of the thread in Fig. \ref{fig:4b}, the situation seems more complicated in that $\epsilon_2/w$ depends on not only the flow rate ratio $\phi$ but also on other parameters such as the viscosity ratio $\chi$. For the same $\phi$, the lower viscosity ratio $\chi$ decreases the major axis and the thread cross section appears more circular.

We also note in Fig. \ref{fig:3} that the cross section of threads become almost stable at $z=2.5$ mm (case 2) and $z=3$ mm (case 3). The long focusing microchannel is necessary for fluids with a large viscosity ratio to produce a thread thin enough. \citet{chung2010numerical} found the thread width $\epsilon_1/w$ also increased with increasing viscosity ratio $\chi$ and predicts the existence of the upper bound of $\chi$ for viscous folding. This is due to the short focusing microchannel, only $2w$ in their study. Consequently, the hydrodynamic focusing procedure is not completely finished and the thread is too thick to undergo viscous folding or buckling instability in the diverging region.

\subsection{Viscous folding}
\label{sec:viscous-folding}

\begin{figure*}
  \centering
  \subfigure[$t=0.178$ s]{
    \includegraphics[width=0.45\textwidth]{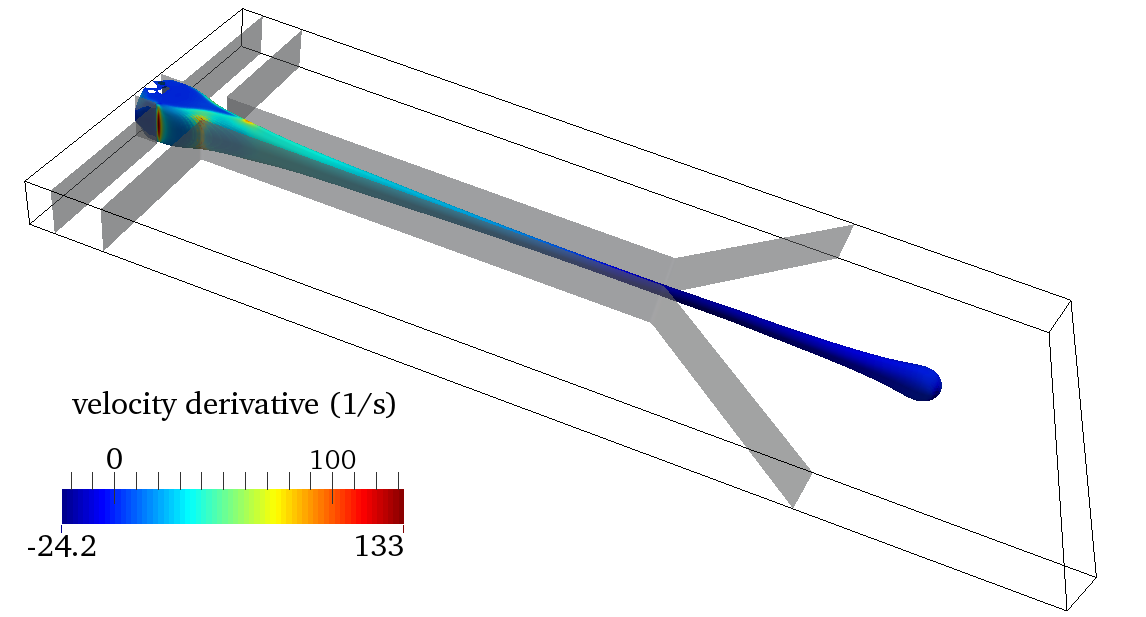}
    \label{fig:5a}
  }
  \subfigure[$t=0.198$ s]{
    \includegraphics[width=0.45\textwidth]{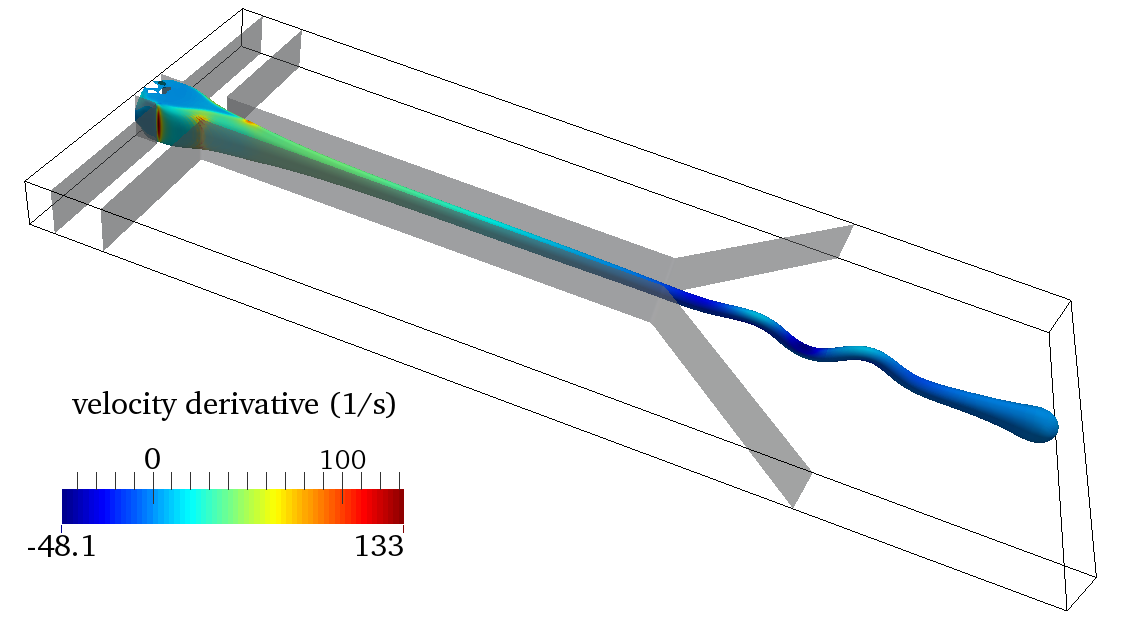}
    \label{fig:5b}
  }
  \subfigure[$t=0.213$ s]{
    \includegraphics[width=0.45\textwidth]{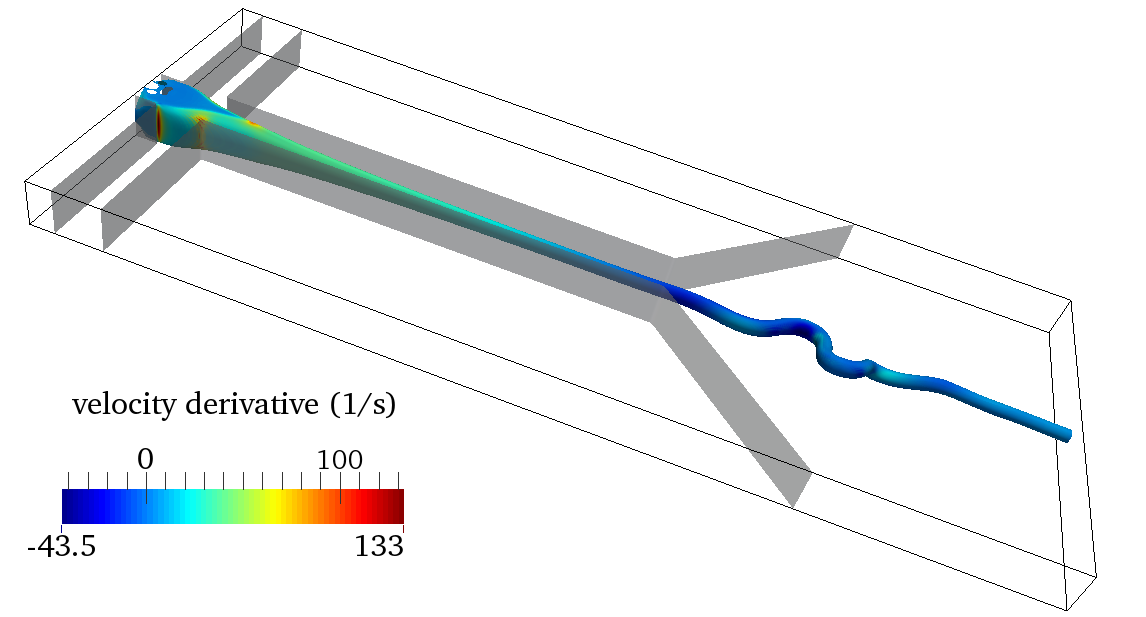}
    \label{fig:5c}
  }
  \subfigure[$t=0.228$ s]{
    \includegraphics[width=0.45\textwidth]{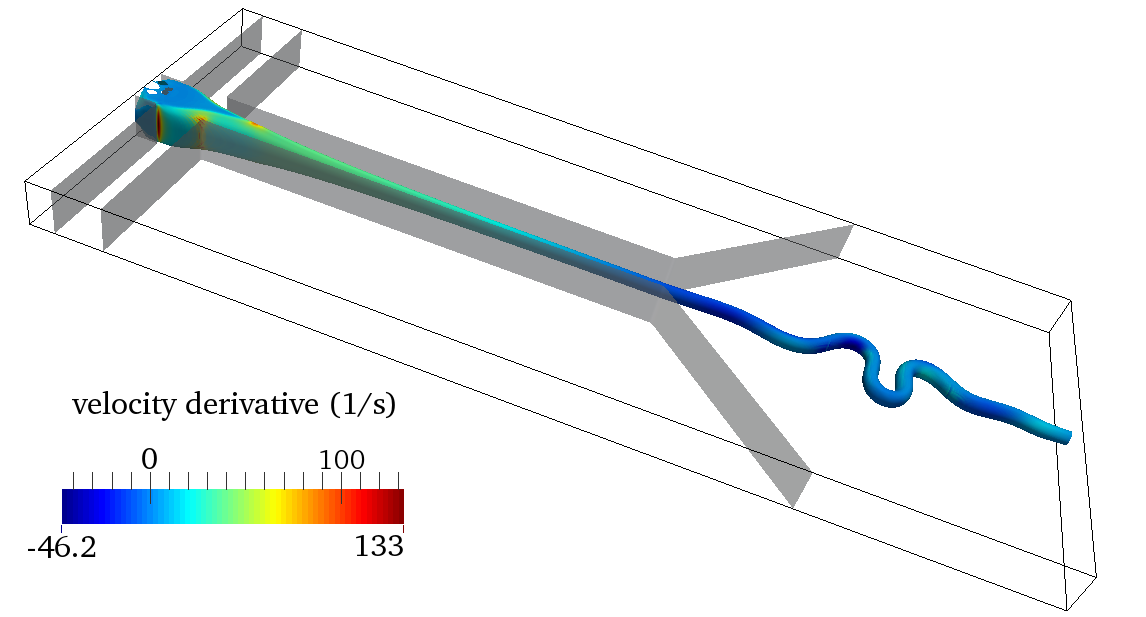}
    \label{fig:5d}
  }
\caption{The flow patterns and contours of the velocity derivative $\partial u_z/ \partial z$ ($\mathrm{s}^{-1}$) at different times for case 1 with $\mathrm{Re}_1=2.74 \times 10^{-4}, \phi=1/12, \chi=2174, \alpha=\pi/2$.}
\label{fig:5}       
\end{figure*}

The thread produced by hydrodynamic focusing continues to flow in the
diverging region and a folding instability appears due to the compressive stress. For our 5 simulation cases, different flow patterns has been observed. In the reference case 1 with $\mathrm{Re}_1=2.74 \times 10^{-4}, \phi=1/12, \chi=2174, \alpha=\pi/2$ as shown in Fig. \ref{fig:5}, the thread begins to fold about an axis in the $y$-direction in Fig. \ref{fig:5b}, and then the folding plane rotates in Fig. \ref{fig:5c}. In Fig. \ref{fig:5d} the new folds appear mainly in the $x$-direction. For case 5 in which only the diverging angle is changed from $\pi/2$ to $\pi$, the flow pattern is similar to the reference case 1.

\begin{figure*}
  \centering
  \subfigure[$t=0.059$ s]{
    \includegraphics[width=0.45\textwidth]{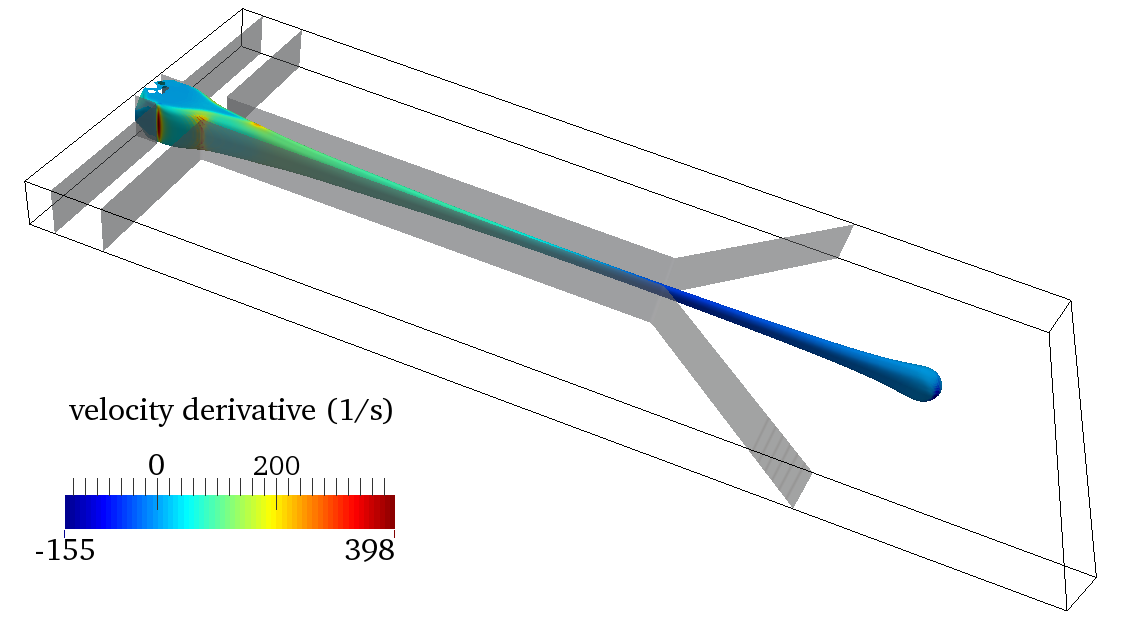}
    \label{fig:6a}
  }
  \subfigure[$t=0.066$ s]{
    \includegraphics[width=0.45\textwidth]{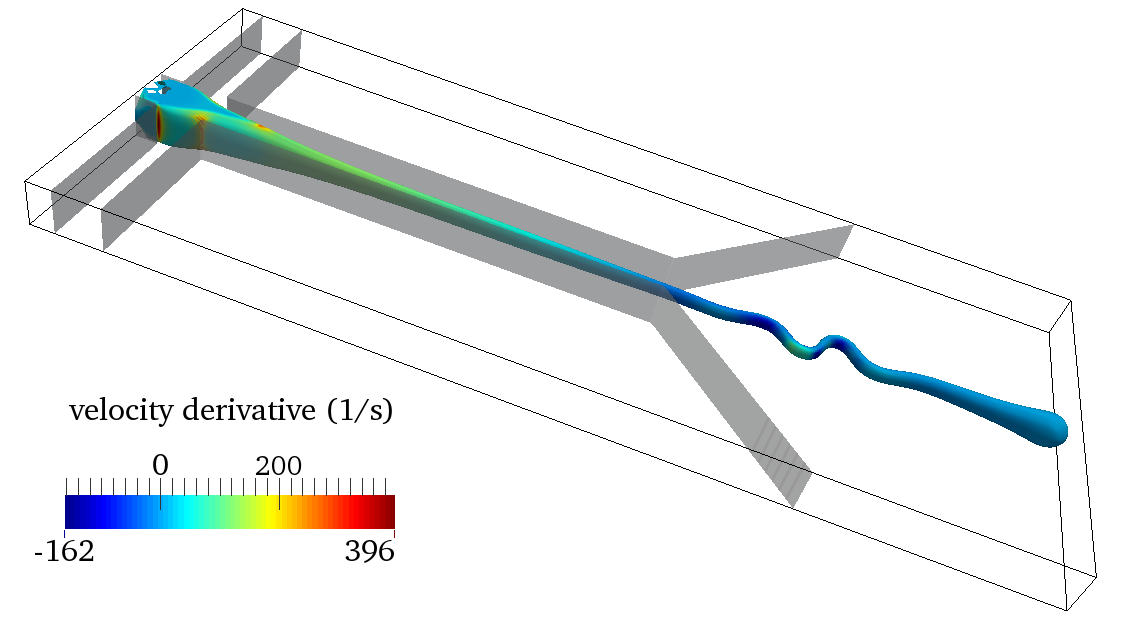}
    \label{fig:6b}
  }
  \subfigure[$t=0.079$ s]{
    \includegraphics[width=0.45\textwidth]{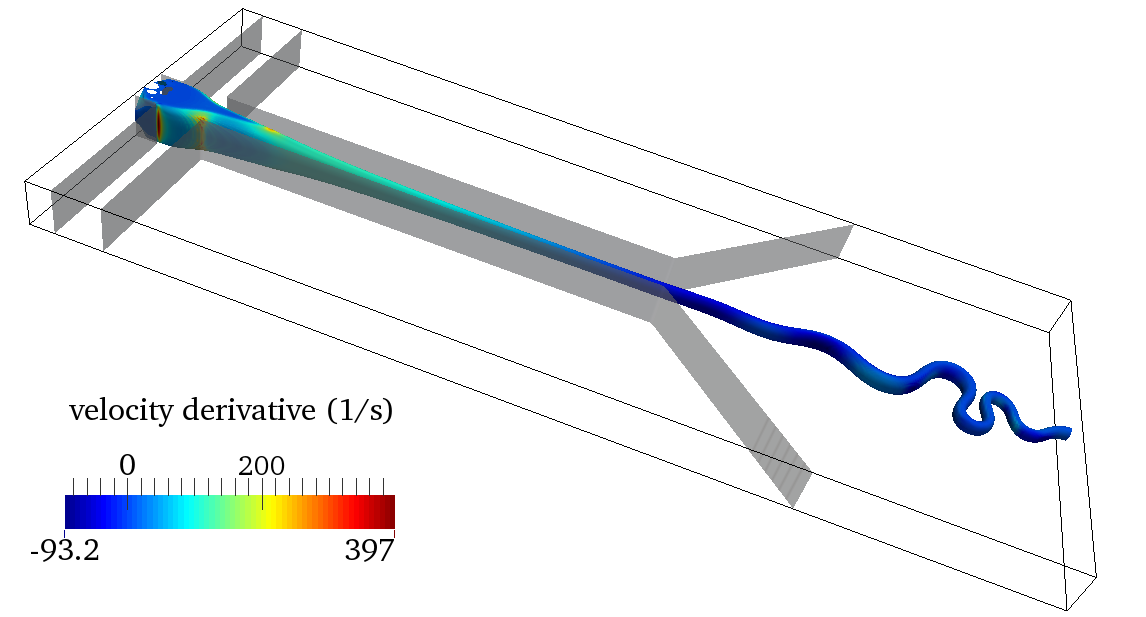}
    \label{fig:6c}
  }
  \subfigure[$t=0.085$ s]{
    \includegraphics[width=0.45\textwidth]{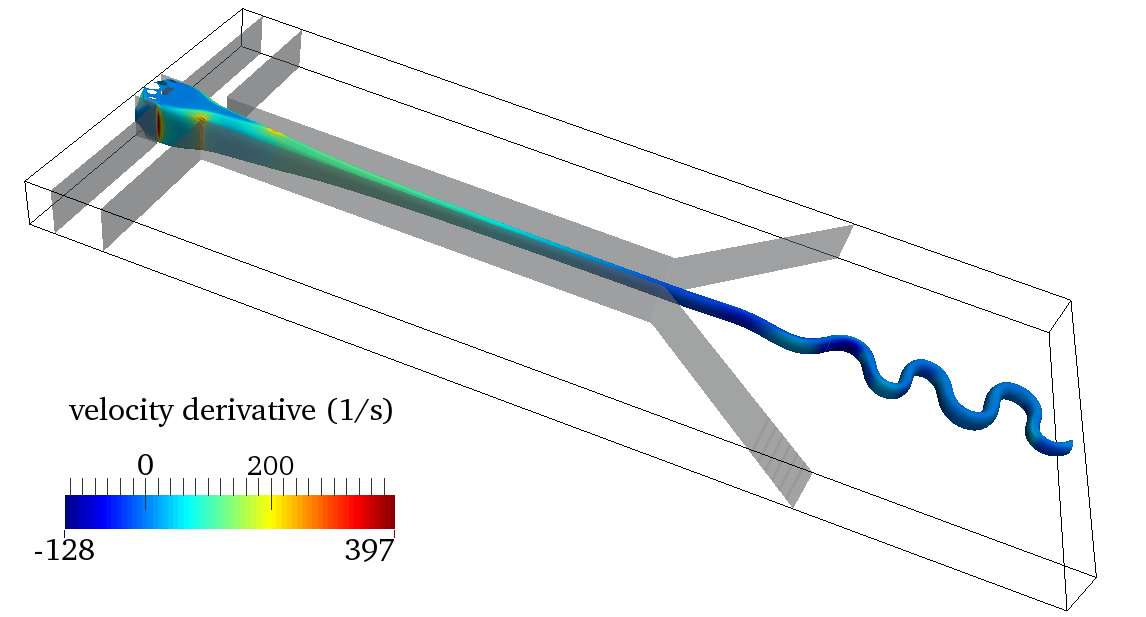}
    \label{fig:6d}
  }
  \subfigure[$t=0.106$ s]{
    \includegraphics[width=0.45\textwidth]{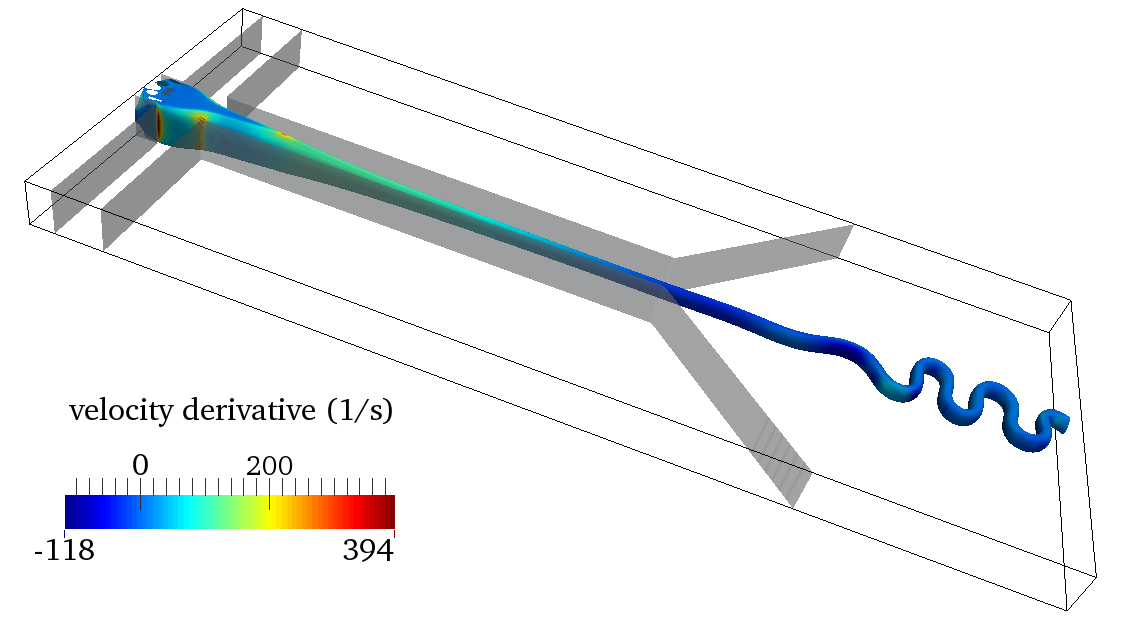}
    \label{fig:6e}
  }
  \subfigure[$t=0.146$ s]{
    \includegraphics[width=0.45\textwidth]{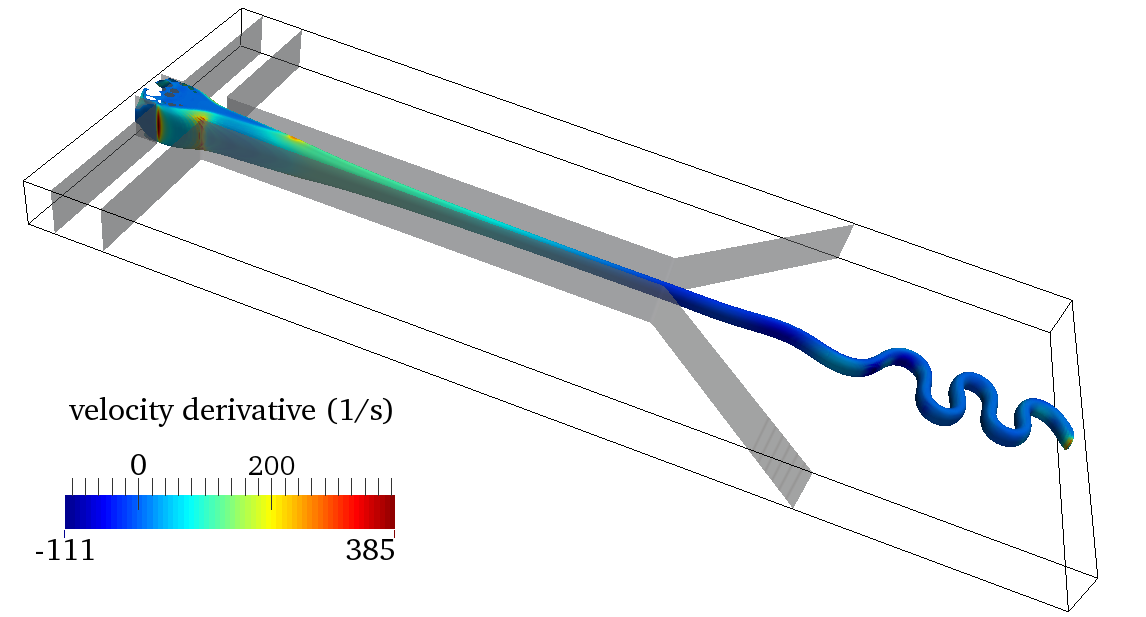}
    \label{fig:6f}
  }
\caption{The flow patterns and contours of the velocity derivative $\partial u_z/ \partial z$ ($\mathrm{s}^{-1}$) at different times for case 2 with $\mathrm{Re}_1=1.64 \times 10^{-4}, \phi=1/12, \chi=2174, \alpha=\pi/2$.}
\label{fig:6}       
\end{figure*}

For simulation case 2 ($\mathrm{Re}_1=1.64 \times 10^{-3}, \phi=1/12, \chi=2174, \alpha=\pi/2$), as shown in Fig. \ref{fig:6a}, the thread begins to fold in the $x$-direction. The folding frequency and amplitude then vary slightly after the thread exits the computation domain in Fig. \ref{fig:6b}, and finally, the folding frequency and amplitude become stable in Fig. \ref{fig:6d} and \ref{fig:6e}. It is noted that the folding only happens in the $x$-direction in simulation case 2.

\begin{figure*}
  \centering
  \subfigure[$t=0.144$ s]{
    \includegraphics[width=0.45\textwidth]{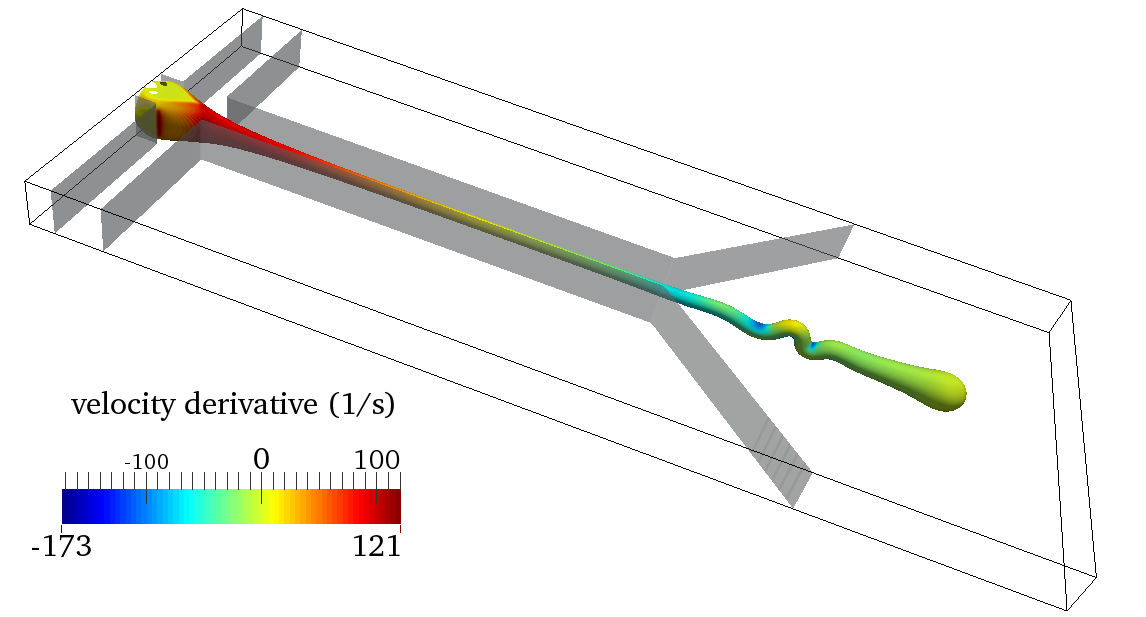}
    \label{fig:7a}
  }
  \subfigure[$t=0.181$ s]{
    \includegraphics[width=0.45\textwidth]{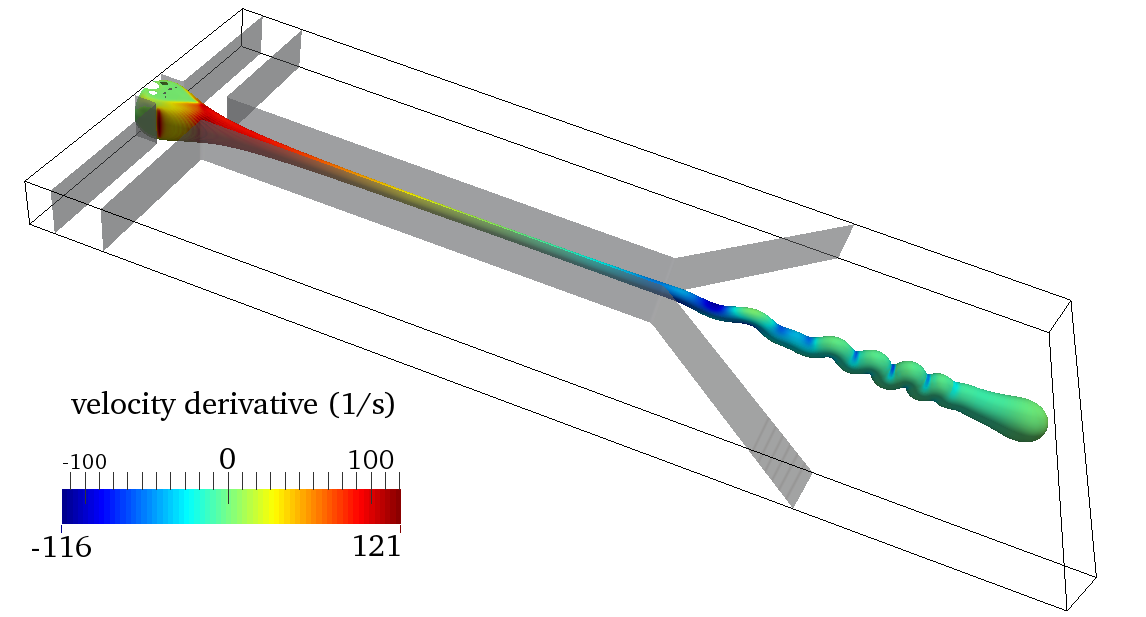}
    \label{fig:7b}
  }
  \subfigure[$t=0.219$ s]{
    \includegraphics[width=0.45\textwidth]{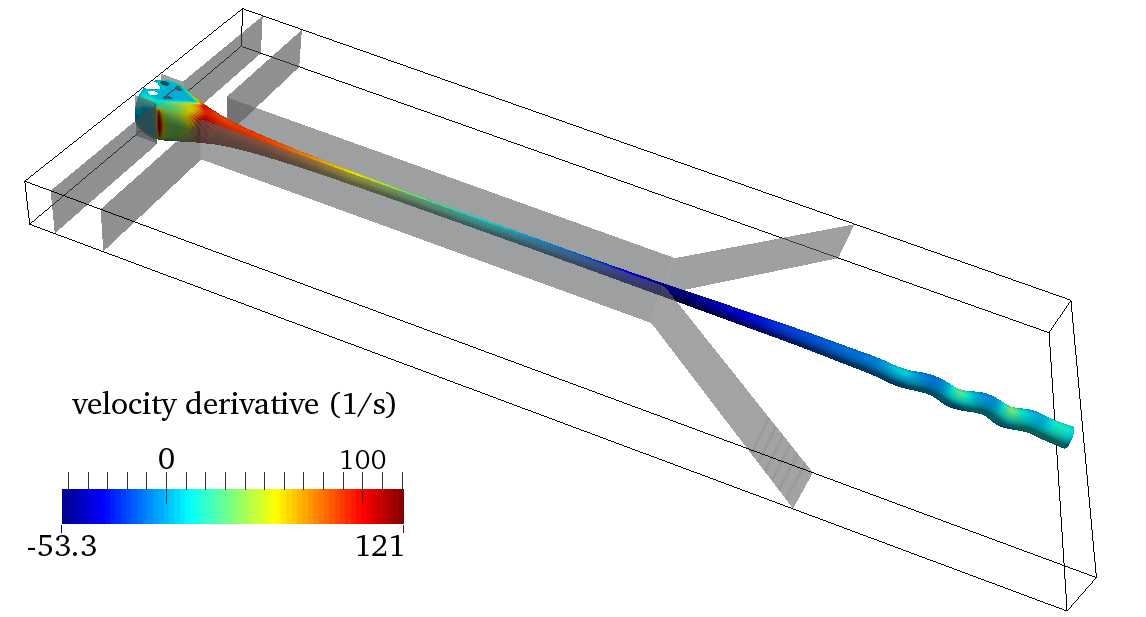}
    \label{fig:7c}
  }
  \subfigure[$t=0.239$ s]{
    \includegraphics[width=0.45\textwidth]{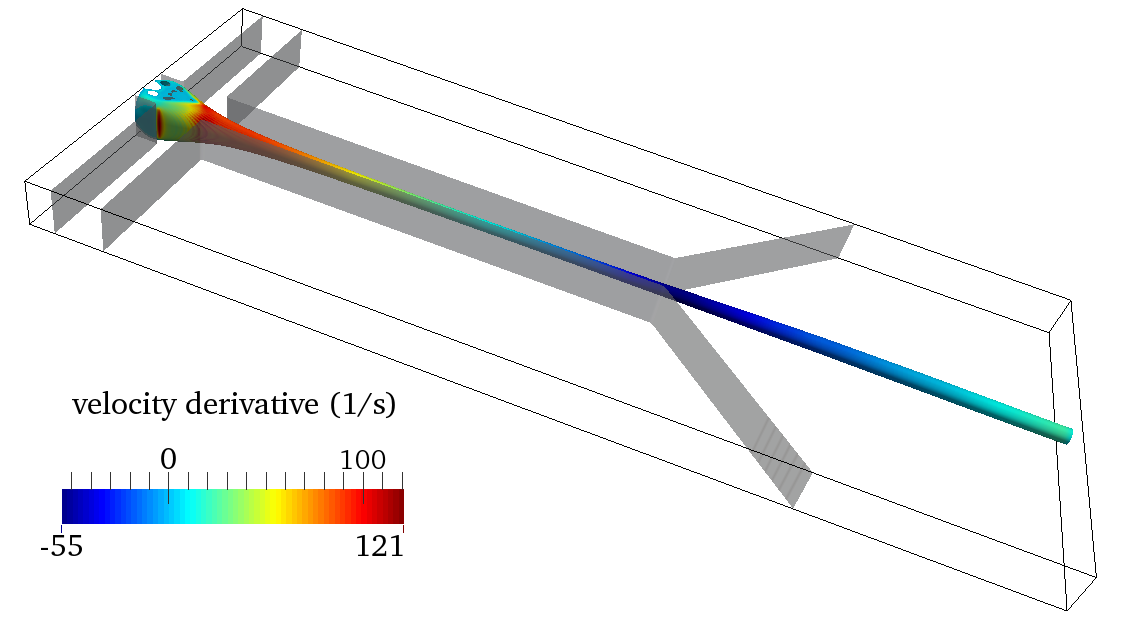}
    \label{fig:7d}
}
\caption{The flow patterns and contours of the velocity derivative $\partial u_z/ \partial z$ ($\mathrm{s}^{-1}$) at different times for case 3 with $\mathrm{Re}_1=2.74 \times 10^{-4}, \phi=1/12, \chi=1000, \alpha=\pi/2$.}
\label{fig:7}       
\end{figure*}

In case 3 with $\mathrm{Re}_1=2.74 \times 10^{-4}, \phi=1/12, \chi=1000, \alpha=\pi/2$, 
the onset of folding appears in the $y$-direction in Fig. \ref{fig:7b}. For this case,
there is not only folding instability but also strong shrinking when the
thread is subject to the compressive stress. The thread is squeezed, so that
the thread becomes fatter and the folding wavelength decreases as the
thread flows downstream (from Fig. \ref{fig:7b} to \ref{fig:7c}). Consequently,
the amplitude of newly appearing folds decreases to zero slowly and its
wavelength becomes larger. Finally, the folding phenomenon disappears
and the thread is completely straight in Fig. \ref{fig:7d}.

\begin{figure*}[!t]
  \centering
  \subfigure[case 1]{
    \includegraphics[width=0.45\textwidth]{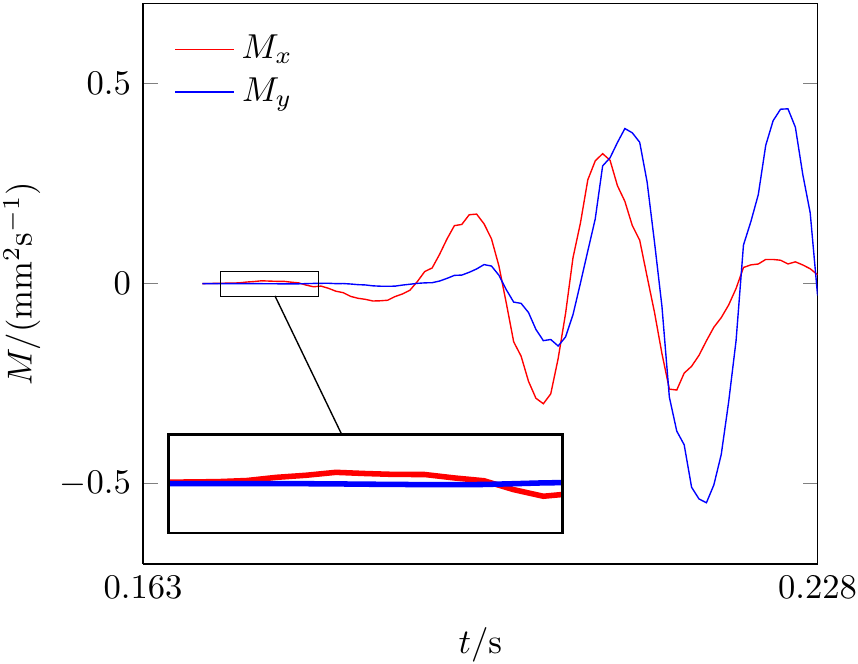}
    \label{fig:8a}
  }
  \subfigure[case 2]{
    \includegraphics[width=0.45\textwidth]{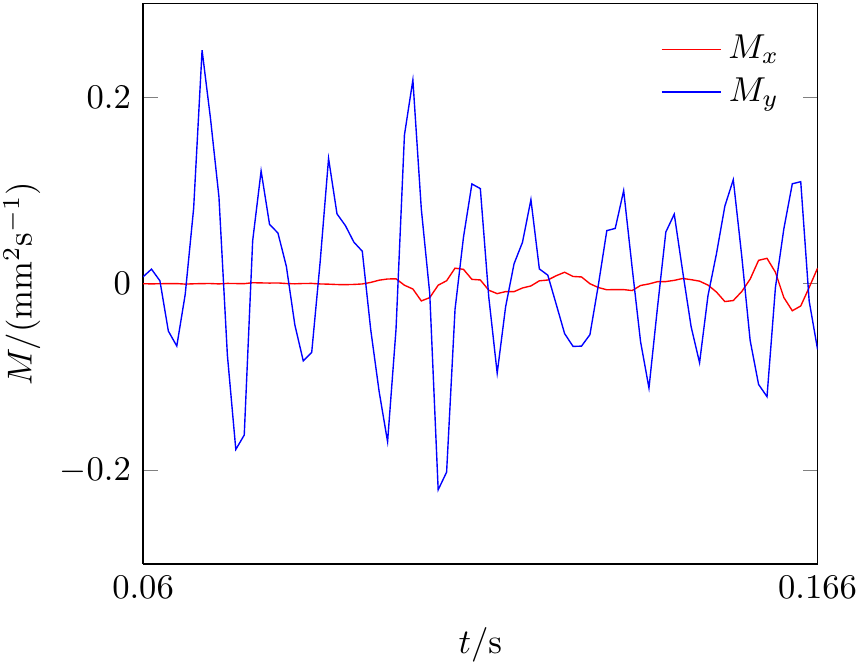}
    \label{fig:8b}
  }
\caption{The bending moment at z = 1.7 mm for case 1 with Re$_1=2.74\times10^{-4}, \phi=1/12, \chi=2174, \alpha=\pi/2$ and case 2 with Re$_1=1.64\times10^{-3}, \phi=1/12, \chi=2174, \alpha=\pi/2$ when the folding instability
occurs.}
\label{fig:8}       
\end{figure*}

The folding is induced by the viscous compressional stress. The velocity of the flow in the long focusing channel and near the diverging point is nearly in the z-direction, i.e. $\boldsymbol{u} = (0, 0, u_z )$. Thus the non-zero components in the viscous stress are
\begin{align}
  \label{eq:5-9} \sigma_{xz}&=\frac{1}{2}\eta_i\frac{\partial u_z}{\partial x}, \\
  \label{eq:5-10} \sigma_{yz}&=\frac{1}{2}\eta_i\frac{\partial u_z}{\partial y}, \\
  \label{eq:5-11} \sigma_{zz}&=-p+\eta_i\frac{\partial u_z}{\partial z},
\end{align}
where $\eta_i$ is the viscosity of liquid $L_1$ or $L_2$. On the cross section of the thread the viscous stress is longitudinal stress, $\sigma_{xz}=\sigma_{yz}=0$ due to the plug flow. In Chung's study \citep{chung2010numerical}, the longitudinal stress is defined as $2\eta_i \partial u_y/ \partial y$ along the centerline. In their Fig. 4(d) \citep{chung2010numerical}, the longitudinal stress is highly compressional. Here, our simulations are 3-dimensional, the longitudinal stress is proportional to the derivatives $\partial u_z/ \partial z$. The derivatives $\partial u_z/ \partial z$ of the velocity $u_z$ with respect to $z$ along the thread are shown in Fig. \ref{fig:8}, it is clear the longitudinal stress is compressional in the diverging region, especially near the diverging point.

On the thread interface, the viscous force per unit area by liquid $L_2$ can be obtained by $\boldsymbol{\sigma} \cdot \boldsymbol{n}$, where $\boldsymbol{n}$ is the unit normal vector to the interface. Since the major axis and minor axis become stable near the diverging point, the unit normal vector is in the $x$-$y$ plane $\boldsymbol{n}=(n_x, n_y, 0)$. Thus, the viscous force per unit area on the interface is 
\begin{equation}
  \label{eq:5-12}
  \boldsymbol{f}_{in}=\boldsymbol{\sigma} \cdot \boldsymbol{n}=(0,0,\sigma_{xz} n_x+\sigma_{yz} n_y)=\frac{1}{2}\eta_2(0,0,\frac{\partial u_z}{\partial n}).
\end{equation}
The viscous force on the interface is proportional to the normal derivative $\partial u_z/\partial n$. Then the bending moment on the cross section of the thread induced by the viscous force on the interface can be calculated, it has two components 
\begin{align}
  \label{eq:5-13}
  \omega_x&=\frac{1}{2}\eta_2 \int_{C} \frac{\partial u_z}{\partial n} (y(s)-y_c)\mathrm{d}s,\\
  \label{eq:5-14}
  \omega_y&=\frac{1}{2}\eta_2 \int_{C} \frac{\partial u_z}{\partial n} (x(s)-x_c)\mathrm{d}s.
\end{align}
Where the integrals are done along the bounding line of cross section $C$, $x_c, y_c$ are the coordinates of the center on the cross section. Here the bending moment is presented by the integral part, i.e. $M_x=2\omega_x/ \eta_2$ and $M_y=2\omega_y/ \eta_2$. For case 1 with Re$_1=2.74\times10^{-4}, \phi=1/12, \chi=2174, \alpha=\pi/2$, the bending moments of the thread $M_x$ and $M_y$ on the cross section at $z=1.7$ mm are plotted from the onset of the folding instability in Fig. \ref{fig:8a}. At first the moment $M_x$ dominates, the cross section rotates about the $x$-axis resulting in folding in the $y$-direction. And then the moment $M_y$ increases, the folding slowly transforms via twisting to folding in the $x$-direction. When the ratio $\epsilon_1/\epsilon_2$ of the thread is much less than $1$ the moment $M_x$ is always very small compared to $M_y$, so that the folding only appears in the $x$-direction. This is just what we observe in simulation case 2 (similar bending moments over time are presented in Fig. \ref{fig:8b}).

Due to the high viscosity contrast and very low Reynolds numbers involved, direct numerical simulations of this two-phase flow are very challenging and to our knowledge these are the first three-dimensional direct parallel numerical simulations of viscous threads in microchannels. However, since the computational time for these simulations is quite long, especially for such viscous threads, the simulations present only the early time onset of the buckling instability of the threads, thus long-time comparisons with experiments for quantities such as folding amplitude and frequency are limited.

\section{Conclusions}
\label{sec:conclusions}

The parallel code BLUE for multi-phase flows was used to simulate three-dimensional viscous folding in diverging microchannels. Liquid $L_1$ takes the form of a thin filament due to hydrodynamic focusing in the long channel that leads to the diverging region. The thread becomes unstable to a folding instability after its entry into the main chamber, due to the longitudinal compressive stress  applied to it by the diverging flow of liquid $L_2$. Given the long computation time for such a low Reynolds number flow, we were limited to a parameter study comprising five simulations in which the flow rate ratio, the viscosity ratio, the Reynolds number, and the shape of the channel were varied relative to a reference model.

During the hydrodynamic focusing, the shape and velocity of the
thread vary dramatically at first, then evolve slowly and finally achieve a
nearly stable state, which implies that the hydrodynamic focusing phase
is complete in the sufficiently long focusing channel. Moreover, the cross section of the thread is elliptical rather than circular. There is a power law relation between the dimensionless minor axis $\epsilon_1/w$ and the flow ratio $\phi$ and our results are in good agreement with experimental and theoretical predictions of other researchers. For the major axis $\epsilon_2$, the situation is more complicated. The lower viscosity ratio $\chi$ decreases the major axis and the thread cross section appears more circular. Additionally, the interfacial tension plays important role in the thread formation after the liquid $L_1$ detaches from walls. Future study will be undertaken to understand the role of interfacial tension on the major axis of the thread produced by hydrodynamic focusing and the following viscous folding instability. 

Unlike the previous two-dimensional simulations of \citet{chung2010numerical}, our simulations are fully three-dimensional and thus do not constrain the axis along which the folding instability could occur. The initial folding axis can be either parallel or perpendicular to the narrow dimension of the chamber. In the former case, the folding slowly transforms via twisting to perpendicular folding, or may remain parallel. The direction of folding onset is determined by the velocity profile and the ellipticity of the thread cross section in the channel that feeds the diverging part of the cell. 

Due to the high viscosity contrast and very low Reynolds numbers involved, direct numerical simulations of this two-phase flow are very challenging and to our knowledge these are the first three-dimensional direct parallel numerical simulations of viscous threads in microchannels. However, since the computational time for these simulations is quite long, especially for such viscous threads, the simulations present only the early time onset of the buckling instability of the threads, thus long-time comparisons with experiments for quantities such as folding amplitude and frequency are limited.
In the future, more long-time simulations with a larger range of viscosity ratio, Reynolds number, flow rate ratio and with different channel geometries will be implemented in order that exhaustive comparisons with experiments can help in improving the understanding of viscous folding in microchannels.

\begin{acknowledgements}
We thank N. Ribe and T. Cubaud for helpful discussions. 
This work was performed using high performance computing resources provided by
the Institut du Developpement et des Ressources en Informatique Scientifique (IDRIS) of the Centre National de la Recherche Scientifique (CNRS).  This research was supported by the Basic Science Research Program through the National Research Foundation of Korea (NRF) funded by the Ministry of Science, ICT and future planning (NRF-2014R1A2A1A11051346)
\end{acknowledgements}

\bibliographystyle{spbasic}      
\bibliography{ref-folding}

\begin{thebibliography}{47}
\providecommand{\natexlab}[1]{#1}
\providecommand{\url}[1]{{#1}}
\providecommand{\urlprefix}{URL }
\expandafter\ifx\csname urlstyle\endcsname\relax
  \providecommand{\doi}[1]{DOI~\discretionary{}{}{}#1}\else
  \providecommand{\doi}{DOI~\discretionary{}{}{}\begingroup
  \urlstyle{rm}\Url}\fi
\providecommand{\eprint}[2][]{\url{#2}}

\bibitem[{Bottausci et~al(2004)Bottausci, Mezi{\'c}, Meinhart, and
  Cardonne}]{bottausci2004mixing}
Bottausci F, Mezi{\'c} I, Meinhart CD, Cardonne C (2004) Mixing in the shear
  superposition micromixer: three-dimensional analysis. Philosophical
  Transactions of the Royal Society of London A: Mathematical, Physical and
  Engineering Sciences 362(1818):1001--1018

\bibitem[{Bringer et~al(2004)Bringer, Gerdts, Song, Tice, and
  Ismagilov}]{bringer2004microfluidic}
Bringer MR, Gerdts CJ, Song H, Tice JD, Ismagilov RF (2004) Microfluidic
  systems for chemical kinetics that rely on chaotic mixing in droplets.
  Philosophical Transactions of the Royal Society of London A: Mathematical,
  Physical and Engineering Sciences 362(1818):1087--1104

\bibitem[{Chang et~al(2007)Chang, Huang, and Yang}]{chang2007three}
Chang CC, Huang ZX, Yang RJ (2007) Three-dimensional hydrodynamic focusing in
  two-layer polydimethylsiloxane (pdms) microchannels. Journal of
  Micromechanics and Microengineering 17(8):1479

\bibitem[{Chen et~al(2009)Chen, Bown, O’Sullivan, MacInnes, Allen, Mulder,
  Blom, and van’t Oever}]{chen2009performance}
Chen Z, Bown M, O’Sullivan B, MacInnes J, Allen R, Mulder M, Blom M, van’t
  Oever R (2009) Performance analysis of a folding flow micromixer.
  Microfluidics and nanofluidics 6(6):763--774

\bibitem[{Chorin(1968)}]{chorin1968numerical}
Chorin AJ (1968) Numerical solution of the navier-stokes equations. Mathematics
  of computation 22(104):745--762

\bibitem[{Chou et~al(2001)Chou, Unger, and Quake}]{chou2001microfabricated}
Chou HP, Unger MA, Quake SR (2001) A microfabricated rotary pump. Biomedical
  Microdevices 3(4):323--330

\bibitem[{Chung et~al(2010)Chung, Choi, Kim, Ahn, and Lee}]{chung2010numerical}
Chung C, Choi D, Kim JM, Ahn KH, Lee SJ (2010) Numerical and experimental
  studies on the viscous folding in diverging microchannels. Microfluidics and
  Nanofluidics 8(6):767--776

\bibitem[{Cruickshank(1988)}]{cruickshank1988low}
Cruickshank J (1988) Low-reynolds-number instabilities in stagnating jet flows.
  Journal of fluid mechanics 193:111--127

\bibitem[{Cruickshank and Munson(1982{\natexlab{a}})}]{cruickshank1982energy}
Cruickshank J, Munson B (1982{\natexlab{a}}) An energy loss coefficient in
  fluid buckling. Physics of Fluids (1958-1988) 25(11):1935--1937

\bibitem[{Cruickshank and Munson(1982{\natexlab{b}})}]{cruickshank1982viscous}
Cruickshank J, Munson B (1982{\natexlab{b}}) The viscous-gravity jet in
  stagnation flow. Journal of Fluids Engineering 104(3):360--362

\bibitem[{Cruickshank and Munson(1983)}]{cruickshank1983theoretical}
Cruickshank J, Munson B (1983) A theoretical prediction of the fluid buckling
  frequency. Physics of Fluids (1958-1988) 26(4):928--930

\bibitem[{Cubaud and Mason(2009)}]{cubaud2009high}
Cubaud T, Mason T (2009) High-viscosity fluid threads in weakly diffusive
  microfluidic systems. New Journal of Physics 11(7):075,029

\bibitem[{Cubaud and Mason(2006{\natexlab{a}})}]{Cubaud2006afolding}
Cubaud T, Mason TG (2006{\natexlab{a}}) Folding of viscous threads in diverging
  microchannels. Physical review letters 96(11):114,501

\bibitem[{Cubaud and Mason(2006{\natexlab{b}})}]{cubaud2006bfolding}
Cubaud T, Mason TG (2006{\natexlab{b}}) Folding of viscous threads in
  microfluidics. Physics of Fluids 18(9):091,108

\bibitem[{Cubaud et~al(2005)Cubaud, Tatineni, Zhong, and Ho}]{cubaud2005bubble}
Cubaud T, Tatineni M, Zhong X, Ho CM (2005) Bubble dispenser in microfluidic
  devices. Physical Review E 72(3):037,302

\bibitem[{Griffiths and Turner(1988)}]{griffiths1988folding}
Griffiths R, Turner J (1988) Folding of viscous plumes impinging on a density
  or viscosity interface. Geophysical Journal International 95(2):397--419

\bibitem[{Habibi et~al(2014)Habibi, Hosseini, Khatami, and
  Ribe}]{habibi2014liquid}
Habibi M, Hosseini S, Khatami M, Ribe N (2014) Liquid supercoiling. Physics of
  Fluids (1994-present) 26(2):024,101

\bibitem[{Harlow et~al(1965)Harlow, Welch et~al}]{harlow1965numerical}
Harlow FH, Welch JE, et~al (1965) Numerical calculation of time-dependent
  viscous incompressible flow of fluid with free surface. Physics of fluids
  8(12):2182

\bibitem[{Joseph and Renardy(1993)}]{joseph2013fundamentals}
Joseph DD, Renardy Y (1993) Fundamentals of two-fluid dynamics: Part II:
  Lubricated Transport, Drops and Miscible Liquids. Springer Science \&
  Business Media

\bibitem[{Kang et~al(2007{\natexlab{a}})Kang, Hulsen, Anderson, den Toonder,
  and Meijer}]{kang2007achaotic}
Kang TG, Hulsen MA, Anderson PD, den Toonder JM, Meijer HE (2007{\natexlab{a}})
  Chaotic advection using passive and externally actuated particles in a
  serpentine channel flow. Chemical Engineering Science 62(23):6677--6686

\bibitem[{Kang et~al(2007{\natexlab{b}})Kang, Hulsen, Anderson, Toonder, and
  Meijer}]{kang2007bchaotic}
Kang TG, Hulsen MA, Anderson PD, Toonder JMd, Meijer HE (2007{\natexlab{b}})
  Chaotic mixing induced by a magnetic chain in a rotating magnetic field.
  Physical Review-Section E-Statistical Nonlinear and Soft Matter Physics
  76(6):66,303--66,303

\bibitem[{Kwak and Lee(2003)}]{kwak2003multigrid}
Kwak DY, Lee JS (2003) Multigrid algorithm for cell-centred finite difference
  method ii : Discontinuous coefficient case. Department of Mathematics, KAIST,
  Taejon Korea pp 305--701

\bibitem[{Lee et~al(2006)Lee, Chang, Huang, and Yang}]{lee2006hydrodynamic}
Lee GB, Chang CC, Huang SB, Yang RJ (2006) The hydrodynamic focusing effect
  inside rectangular microchannels. Journal of Micromechanics and
  Microengineering 16(5):1024

\bibitem[{Mahadevan et~al(1998)Mahadevan, Ryu, and Samuel}]{mahadevan1998fluid}
Mahadevan L, Ryu WS, Samuel AD (1998) Fluid'rope trick'investigated. Nature
  392(6672):140--140

\bibitem[{Maleki et~al(2004)Maleki, Habibi, Golestanian, Ribe, and
  Bonn}]{maleki2004liquid}
Maleki M, Habibi M, Golestanian R, Ribe N, Bonn D (2004) Liquid rope coiling on
  a solid surface. Physical review letters 93(21):214,502

\bibitem[{Meleson et~al(2004)Meleson, Graves, and Mason}]{meleson2004formation}
Meleson K, Graves S, Mason TG (2004) Formation of concentrated nanoemulsions by
  extreme shear. Soft Materials 2(2-3):109--123

\bibitem[{Paik et~al(2003{\natexlab{a}})Paik, Pamula, and Fair}]{paik2003rapid}
Paik P, Pamula VK, Fair RB (2003{\natexlab{a}}) Rapid droplet mixers for
  digital microfluidic systems. Lab on a Chip 3(4):253--259

\bibitem[{Paik et~al(2003{\natexlab{b}})Paik, Pamula, Pollack, and
  Fair}]{paik2003electrowetting}
Paik P, Pamula VK, Pollack MG, Fair RB (2003{\natexlab{b}})
  Electrowetting-based droplet mixers for microfluidic systems. Lab on a Chip
  3(1):28--33

\bibitem[{Peskin(1977)}]{peskin1977numerical}
Peskin CS (1977) Numerical analysis of blood flow in the heart. Journal of
  computational physics 25(3):220--252

\bibitem[{Pollack et~al(2002)Pollack, Shenderov, and
  Fair}]{pollack2002electrowetting}
Pollack M, Shenderov A, Fair R (2002) Electrowetting-based actuation of
  droplets for integrated microfluidics. Lab on a Chip 2(2):96--101

\bibitem[{Ribe(2004)}]{ribe2004coiling}
Ribe NM (2004) Coiling of viscous jets. In: Proceedings of the Royal Society of
  London A: Mathematical, Physical and Engineering Sciences, vol 460, pp
  3223--3239

\bibitem[{Ribe et~al(2006)Ribe, Lister, and Chiu-Webster}]{ribe2006stability}
Ribe NM, Lister JR, Chiu-Webster S (2006) Stability of a dragged viscous
  thread: Onset of “stitching” in a fluid-mechanical “sewing machine”.
  Physics of Fluids 18(12):124,105

\bibitem[{Rida and Gijs(2004)}]{rida2004manipulation}
Rida A, Gijs M (2004) Manipulation of self-assembled structures of magnetic
  beads for microfluidic mixing and assaying. Analytical chemistry
  76(21):6239--6246

\bibitem[{Shin and Juric(2007)}]{shin2007high}
Shin S, Juric D (2007) High order level contour reconstruction method. Journal
  of mechanical science and technology 21(2):311--326

\bibitem[{Shin and Juric(2009{\natexlab{a}})}]{shin2009hybrid}
Shin S, Juric D (2009{\natexlab{a}}) A hybrid interface method for
  three-dimensional multiphase flows based on front tracking and level set
  techniques. International Journal for Numerical Methods in Fluids
  60(7):753--778

\bibitem[{Shin and Juric(2009{\natexlab{b}})}]{shin2009simulation}
Shin S, Juric D (2009{\natexlab{b}}) Simulation of droplet impact on a solid
  surface using the level contour reconstruction method. Journal of mechanical
  science and technology 23(9):2434--2443

\bibitem[{Shin et~al(2014)Shin, Chergui, and Juric}]{shin2014solver}
Shin S, Chergui J, Juric D (2014) A solver for massively parallel direct
  numerical simulation of three-dimensional multiphase flows. arXiv preprint
  arXiv:14108568

\bibitem[{Simonnet and Groisman(2005)}]{simonnet2005two}
Simonnet C, Groisman A (2005) Two-dimensional hydrodynamic focusing in a simple
  microfluidic device. Applied Physics Letters 87(11):114,104

\bibitem[{Skorobogatiy and Mahadevan(2000)}]{skorobogatiy2000folding}
Skorobogatiy M, Mahadevan L (2000) Folding of viscous sheets and filaments. EPL
  (Europhysics Letters) 52(5):532

\bibitem[{Stiles et~al(2005)Stiles, Fallon, Vestad, Oakey, Marr, Squier, and
  Jimenez}]{stiles2005hydrodynamic}
Stiles T, Fallon R, Vestad T, Oakey J, Marr D, Squier J, Jimenez R (2005)
  Hydrodynamic focusing for vacuum-pumped microfluidics. Microfluidics and
  Nanofluidics 1(3):280--283

\bibitem[{Stroock et~al(2002{\natexlab{a}})Stroock, Dertinger, Ajdari,
  Mezi{\'c}, Stone, and Whitesides}]{stroock2002chaotic}
Stroock AD, Dertinger SK, Ajdari A, Mezi{\'c} I, Stone HA, Whitesides GM
  (2002{\natexlab{a}}) Chaotic mixer for microchannels. Science
  295(5555):647--651

\bibitem[{Stroock et~al(2002{\natexlab{b}})Stroock, Dertinger, Whitesides, and
  Ajdari}]{stroock2002patterning}
Stroock AD, Dertinger SK, Whitesides GM, Ajdari A (2002{\natexlab{b}})
  Patterning flows using grooved surfaces. Analytical Chemistry
  74(20):5306--5312

\bibitem[{Taylor(1969)}]{taylor1969instability}
Taylor G (1969) Instability of jets, threads, and sheets of viscous fluid. In:
  Applied Mechanics, Springer, pp 382--388

\bibitem[{Tchavdarov et~al(1993)Tchavdarov, Yarin, and
  Radev}]{tchavdarov1993buckling}
Tchavdarov B, Yarin A, Radev S (1993) Buckling of thin liquid jets. Journal of
  Fluid Mechanics 253:593--615

\bibitem[{Tome and Mckee(1999)}]{tome1999numerical}
Tome MF, Mckee S (1999) Numerical simulation of viscous flow: buckling of
  planar jets. International journal for numerical methods in fluids
  29(6):705--718

\bibitem[{Utada et~al(2005)Utada, Lorenceau, Link, Kaplan, Stone, and
  Weitz}]{utada2005monodisperse}
Utada A, Lorenceau E, Link D, Kaplan P, Stone H, Weitz D (2005) Monodisperse
  double emulsions generated from a microcapillary device. Science
  308(5721):537--541

\bibitem[{Wu and Nguyen(2005)}]{wu2005hydrodynamic}
Wu Z, Nguyen NT (2005) Hydrodynamic focusing in microchannels under
  consideration of diffusive dispersion: theories and experiments. Sensors and
  Actuators B: Chemical 107(2):965--974

\end{thebibliography}


%
%

\end{document}